\documentclass[pdflatex,sn-aps,iicol]{sn-jnl}

\usepackage{array}
\newcolumntype{C}[1]{>{\centering\arraybackslash}p{#1}}

\usepackage{graphicx}
\usepackage{dcolumn}
\usepackage{bm}
\usepackage{amsmath}
\usepackage[amssymb]{SIunits}

\usepackage{lineno}

\usepackage{multirow}%
\usepackage{amsmath,amssymb,amsfonts}%
\usepackage{amsthm}%
\usepackage{mathrsfs}%
\usepackage[title]{appendix}%
\usepackage{xcolor}%
\usepackage{textcomp}%
\usepackage{manyfoot}%
\usepackage{booktabs}%
\usepackage{algorithm}%
\usepackage{algorithmicx}%
\usepackage{algpseudocode}%
\usepackage{listings}%

\begin{document}

\title{Chiral Particles in Taylor--Couette Turbulence}

\author*[1]{\fnm{Mees M.} \sur{Flapper}}\email{m.m.flapper@utwente.nl}

\author*[1,2]{\fnm{Detlef} \sur{Lohse}}\email{d.lohse@utwente.nl}
\author*[1]{\fnm{Sander G.} \sur{Huisman}}\email{s.g.huisman@utwente.nl}

\affil[1]{\orgdiv{Physics of Fluids department and Max Planck Center for Complex Fluid Dynamics, J. M. Burgers Centre for Fluid Dynamics}, \orgname{University of Twente}, \orgaddress{P.O. Box 217, \postcode{7500NB}, \city{Enschede},  \country{The Netherlands}}}

\affil[2]{\orgdiv{Max Planck Institute for Dynamics and Self-Organization}, \orgaddress{\street{Am Faßberg 17}, \city{Göttingen}, \country{Germany}}}

\abstract{\unboldmath This work investigates chiral particles, which break mirror symmetry, in turbulent Taylor--Couette flow. These particles generally display a translation-rotation coupling moving through a quiescent fluid. Here we performed experiments using large chiral particles (typical size \unit{5}{mm}) in turbulent Taylor--Couette flow, for Reynolds numbers $9\cdot10^3 \leq \text{Re} \leq 1.5 \cdot 10^5$. The density-matched chiral particles are studied in a dilute regime $(\phi = 1.7 \cdot 10^{-4})$, where their location and orientation are tracked over time to investigate the particle-fluid coupling. 
We investigate whether the translation-rotation coupling observed at low Reynolds numbers is still observable over the measured high Reynolds numbers, using the tracked location and orientation. Similarly, we verify whether the chiral particles display a preferred location or orientation, and whether the left-handed and right-handed particles show different rotation statistics. The location data show that the chiral particles closely follow the structure of Taylor vortices. Hence, the orientation data and rotation data of the chiral particles are split between the Taylor vortices and particle chiralities. The results show no difference in rotation and orientation dynamics between chiralities. Rather, the particle dynamics are flow-dominated, where the flow vorticity determines the specific particle dynamics.}

\maketitle

\section{Introduction}

\begin{figure}
    \centering
    \includegraphics[width = 0.45 \textwidth]{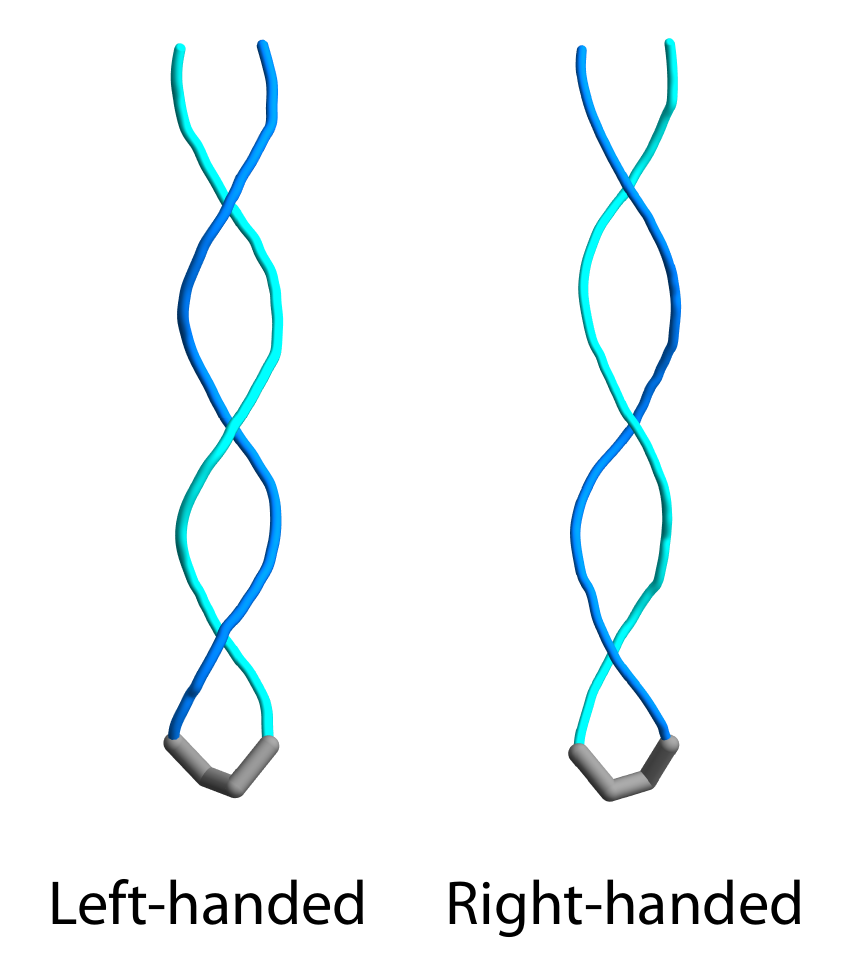}
    \caption{Trajectories of heavy settling chiral particles in quiescent flow for both particle chiralities. The difference in rotation direction is highlighted by the trajectories of the endpoints of the particle. These trajectories are obtained from experimental measurements, performed similarly to the measurements as described in Section \ref{sec:Method and setup}.}
    \label{fig:Settling_chiral_trajectories}
\end{figure}

Fluid flows containing some kind of particles are ubiquitous. Beyond the microscales, water- or airflows are turbulent in nature, where the particles seeding these flows are generally anisotropic, with spherical particles being an idealised exception. Examples of such particle-laden turbulent flows are sediment in rivers (either in the bulk or at the surface) \cite{Vercruysse2017,Salmon2023}, snow falling in air \cite{Nemes2017}, and recycling processes \cite{Bauer2018}. Whereas spherical particles have been thoroughly studied for decades \cite{Li1992,Brown2009,Zimmermann2011,Baroudi2023}, more recently, various anisotropic particles have received more attention \cite{Byron2015,Voth2017,Anand2020,Fries2017,Olivieri2022}. 
\begin{figure*}[t!]
    \centering
    \includegraphics[width=0.9\linewidth]{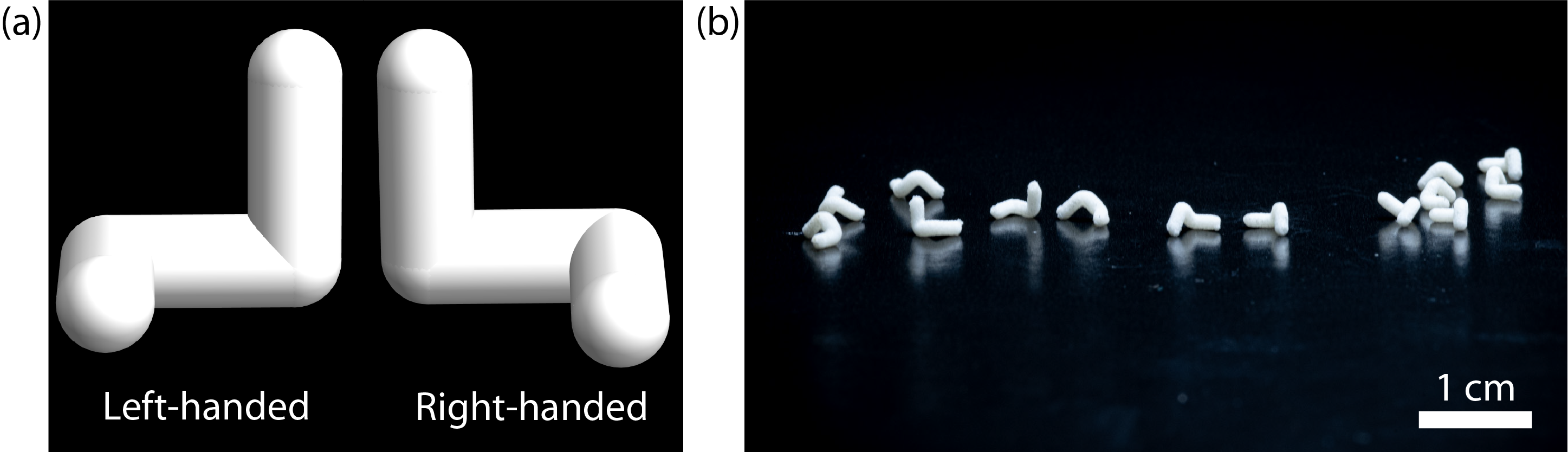}
    \caption{(a) Synthetic 3D representation of the chiral particles used in this research. (b) Photograph of the chiral particles used in the Taylor--Couette facility. Both left-handed and right-handed particles are shown in the image.}
    \label{fig:Particles synthetic}
\end{figure*}
Anisotropic particles can vary in shape and size, giving rise to a wide range of rich particle dynamics: curved sheets display a regular oscillating motion as they settle in a quiescent fluid \cite{Chan2021}, whereas disks settling in quiescent air exhibit multiple falling modes \cite{Tinklenberg2023}. In turbulent channel flow, non-axisymmetric fibers increasingly tend to cluster near the wall as their curvature increases \cite{Alipour2022}. Inertial particles generally tend to cluster in turbulent Taylor--Couette flows, where this clustering is more pronounced at higher Stokes numbers \cite{Jiang2024}. The precise location and pattern in which the particles cluster depends on the type of Taylor--Couette flow \cite{Majji2018}. Besides clustering, anisotropic particles may show preferential orientations in turbulent flows \cite{Parsa2012}. Prolate spheroids display a strong preferential alignment in Taylor--Couette flows and get trapped in the Taylor vortex core for specific Taylor numbers \cite{Assen2022}. Similarly, rodlike fibers can align with the vorticity vector \cite{Pumir2011}, and show a preferential orientation even at high Reynolds numbers \cite{Bakhuis2019}. Expanding on axisymmetric spheroids, jacks and crosses were investigated by Marcus et al. \cite{Marcus2014}, showing that crosses exhibit a preferential orientation, with an arm of the cross aligning with its solid body rotation rate vector. The jacks do not show any preferential orientation and rotate like a sphere. Exploring shapes other than spheroids or axisymmetric particles is done recently for example by Sun et al. \cite{Sun2024}, who demonstrate that triangular, elliptic, and L-shaped particles have multiple stable orientations when settling in a quiescent medium. \\
\\
We now come to the focus of this paper: \textit{chiral} particles. Besides breaking isotropy and axisymmetry, these particles break mirror symmetry in addition, which gives rise to rich particle dynamics. Due to the common occurrence and importance of chiral particles in nature and industry, these particles have already been studied, though mainly on the microscales \cite{Zhong2022}. Different works have shown that these chiral particles can be separated using various flows \cite{Meinhardt2012,Aristov2013}, indicating distinct differences in particle-flow dynamics of the left-handed and right-handed particles. Besides applications in chemical engineering, chiral particles also occur in nature at larger scales. These chiral particles generally have a translation-rotation coupling, as seen in maple seeds \cite{Varshney2012}, where the seeds are autorotating helicopters, increasing the distance they travel. At these larger scales, chiral particles have only been studied recently in fluid mechanics. Kramel et al. \cite{Kramel2016} showed that a chiral dipole has a preferential rotation in isotropic turbulence due to the chiral ends of the particle. This preferential rotation due to chirality was shown again for settling helical ribbons by Huseby et al., where the translation-rotation coupling leads to quasi-periodic angular dynamics and complex spatial trajectories \cite{Huseby2024}.\\
\\
This paper investigates simple chiral particles, similar to particles used by Aristov et al. \cite{Aristov2013}, tracked in turbulent Taylor--Couette flow. A synthetic image of the investigated particles is shown in Figure \ref{fig:Particles synthetic}(a). The right-handed particle is shown on the right, and the left-handed particle on the left. A collection of these 3D-printed chiral particles is shown in Figure \ref{fig:Particles synthetic}(b). These particles display a translation-rotation coupling when settling in a quiescent fluid, as shown by Piumini et al. \cite{Piumini2024}. The chirality of the particle determines its rotation direction through the particle shape, where the left-handed and right-handed particles have opposite rotation vectors when settling. The settling dynamics of these heavy chiral particles in a quiescent fluid are illustrated in Figure \ref{fig:Settling_chiral_trajectories}. These illustrations are made by tracking heavy chiral particles experimentally, similar to the tracking described in Section \ref{subsec: part track}.
Piumini et al. \cite{Piumini2024} then added homogeneous isotropic turbulent forcing, which resulted in more isotropic rotation and orientation dynamics for the chiral particles. Therefore, the preferential rotation and orientation of the chiral particles gradually vanish when increasing the homogeneous isotropic turbulent forcing.

Rather than settling heavy particles in a quiescent fluid (displaying the translation-rotation coupling), this paper investigates the rotation and orientation statistics of chiral particles in turbulent Taylor--Couette flow (more specifically, in turbulent Taylor vortices). The main reason for choosing the Taylor--Couette facility is the regular Taylor vortices it creates, and the well-controlled but turbulent flow. The particle dynamics in Taylor--Couette flow are expected to be vastly different compared to homogeneous isotropic turbulence, due to the different flows: the Taylor--Couette flow is anisotropic and inhomogeneous. Particles in turbulent Taylor--Couette flow experience a strongly sheared flow, resulting in a velocity gradient over the radial direction, whereas any velocity fluctuations over the particle-scale in the simulations by Piumini et al. \cite{Piumini2024} are isotropic. As a result, the driving of the fluid-particle interaction is fundamentally different. Therefore, although the strong turbulent flow may suggest isotropic orientation dynamics as found by Piumini et al. \cite{Piumini2024} for homogeneous isotropic turbulence, the results can starkly differ in Taylor--Couette flow. For example, a clear preferential orientation has been found for finite-size fibers in turbulent Taylor-Couette flow by Bakhuis et al. \cite{Bakhuis2019}, despite a high Reynolds number and particles larger than the Kolmogorov scale. If we find such a preferential orientation, the shear flow and velocity fluctuations may produce a preferential particle rotation due to the particle geometry.

The Taylor--Couette flow is axially symmetric, with the dominant flow velocity in the azimuthal direction, where numerous Taylor vortices are stacked vertically. The vorticity vector of the stacked Taylor vortices alternates between aligned with and opposed to the azimuthal direction as illustrated by Huisman et al \cite{Huisman2015}. Therefore, two opposing vorticity directions are present in the flow simultaneously. This allows us to investigate whether the left-handed and right-handed particles show different orientation or rotation statistics depending on the flow vorticity and particle chirality. Furthermore, we investigate whether the particles can be separated by handedness as a result of the flow vorticity, as seen on smaller scales at lower Reynolds numbers \cite{Meinhardt2012,Aristov2013}. Tracking the orientation of the chiral particles over time will show if any preferential orientation or rotation exists, whereas tracking the particles' location indicates whether the vorticity of the flow separates the particles by handedness.\\
The paper is organised as follows: Section \ref{sec:Method and setup} describes the experimental setup, and the performed measurements. The results are shown in Section \ref{sec:Results}, and we close with conclusions and an outlook in Section \ref{sec: Outlook & Conclusion}.

\section{Methods and Setup} \label{sec:Method and setup}

In order to experimentally investigate the location, orientation, and rotation statistics of the millimetric chiral particles, shown in Figure \ref{fig:Particles synthetic}(b), in Taylor--Couette flow, we tracked the location and orientation of these chiral particles (both left-handed and right-handed). The particles were 3D printed by the company Shapeways using a sintering 3D printing method, resulting in rough, partially porous particles, with a density slightly higher than water. Since the particle-vortex interaction is the main focus of this study, the particle volume fraction is low $(\phi = 1.7 \cdot 10^{-4})$ to keep the particle-particle interactions minimal \cite{Elghobashi1994}. Additionally, the fluid and particles need to be density matched to eliminate any gravity effects. Therefore, the fluid was density matched to the particles by adding glycerol to the water in the Taylor--Couette setup. This was achieved with a water-glycerol mixture containing 11\% glycerol by mass. 
\begin{figure*}
    \centering
    \includegraphics[width = 0.8\textwidth]{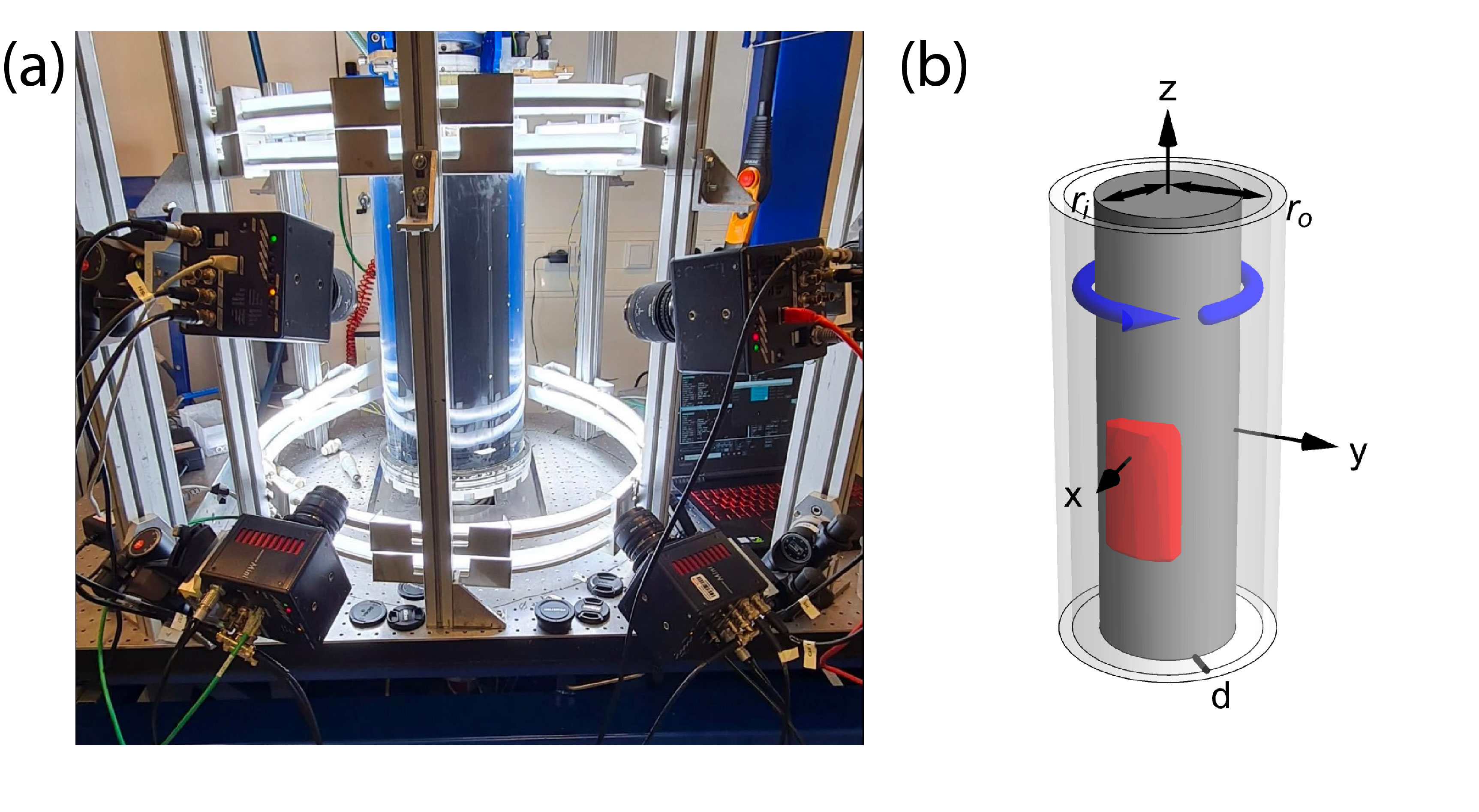}
    \caption{(a) Image of the used Taylor--Couette setup with 4 high-speed cameras used to image particles at (approximately) mid-height of the setup. (b) Schematic of the Taylor--Couette setup and the coordinate system in the lab frame. The measurement volume is shown in red, the blue arrow shows the direction of the dominant flow.}
    \label{fig:Setup_schematic}
\end{figure*}
The experiments were performed in the Boiling Twente Taylor--Couette Facility \cite{Huisman2015} shown in Figure \ref{fig:Setup_schematic}(a), where the water-glycerol mixture is confined between two concentric cylinders. The radii of the inner and outer cylinders are $r_i =$ \unit{75}{mm} and $r_o =$ \unit{105}{mm}, respectively, resulting in a radius ratio $\eta = r_i/r_o = 0.714$, and a gap width $d = r_o - r_i = $ \unit{30}{mm}. The height of the gap is $L = $ \unit{549}{mm}, giving the aspect ratio $\Gamma = L/d = 18.3$. The cylinders counterrotated with a fixed ratio of $a = -f_o/f_i = 0.4$ between the rotation frequencies of the inner and outer cylinder, which is known to create strong Taylor vortices \cite{Gul2017}. The most important dimensions are shown in the schematic setup displayed in Figure \ref{fig:Setup_schematic}(b), which also defines the lab coordinate system. The red volume illustrates the measurement volume, whereas the blue arrow shows the dominant flow velocity. Between the measurement sets, we vary the inner cylinder rotation from $f_i =$ \unit{0.625}{Hz} to $f_i =$ \unit{10}{Hz}, by multiplying by factors of 2. We define the Reynolds number based on the difference in velocities between the cylinders:
\begin{equation}
    \text{Re} = \frac{(\omega_i r_i - \omega_o r_o) (r_o - r_i)}{\nu},
\end{equation}
where $\omega_i$ and $\omega_o$ are the angular velocity of the inner and outer cylinder respectively. The inner and outer cylinder radii are defined as before, and $\nu$ is the kinematic viscosity of the water-glycerol mixture. The kinematic viscosity of the mixture was $\nu = $ \unit{1.3 \cdot 10^{-6}}{\meter^2 \per \second}, using documented properties of glycerol solutions \cite{glycerine1963}. The resulting Reynolds numbers ranged over $9.5 \cdot 10^3 \leq \text{Re} \leq 1.5 \cdot 10^5$, where the dominant flow velocity is in the azimuthal direction. Furthermore, the driving strength of the system is given by the Taylor number, defined as 
\begin{equation}
    \text{Ta} = \frac{(1 + \eta)^4}{64\eta^2} \frac{(r_o - r_i)^2 (r_i + r_o)^2 (\omega_i - \omega_o)^2}{\nu^2},
\end{equation}
in analogy to the Rayleigh number in Rayleigh--Bénard convection \cite{Eckhardt2007}. The characteristic flow and particle parameters are shown in table \ref{tab:flow parameters}. The energy dissipation rate $\epsilon$ is used to compute the Kolmogorov scales, and its value is found from simulations on the same experimental setup \cite{Ostilla2014,Ezeta2018}, where we extracted the relevant data using PlotDigitizer\cite{PlotDigitizer}. Finally, we define the Stokes number as the ratio between the particle relaxation time and the Kolmogorov timescale: 
\begin{equation}
    \text{Stk} = t_0/\tau_\eta,
\end{equation}
where we estimate the particle relaxation time as $t_0 = \frac{\rho_p d_p^2}{18 \mu}$. Here $\rho_p$ is the particle density, $d_p$ is the typical particle size (\unit{5}{mm} in this case), and $\mu$ is the dynamic viscosity of the fluid. The Kolmogorov timescale is defined as $\tau_\eta = \sqrt{\frac{\nu}{\epsilon}}$.
\begin{table*}[]
\begin{tabular}{C{1.3cm} C{1.3cm} C{1.8cm} C{1.8cm} C{1.8cm} C{1.8cm} C{1.8cm} C{1.8cm}}
\hline
$\omega_i/ (2 \pi)$ & $\omega_o / (2 \pi) $ & $\text{Re}$ & Ta &    $\epsilon$ &    $\tau_\eta$ & $\eta_k$ & $\text{Stk}$\\
(1/s)    &   (1/s)  &    &    &  ($\text{m}^2 / \text{s}^3$) &   (s)  &   (m) &  \\ \hline
    $0.625$     &    $-0.25$      &   $9.5 \cdot 10^{3}$    &  $1.4 \cdot 10^{8}$  &  $4.0 \cdot 10^{-3}$  & $1.8 \cdot 10^{-2}$  & $1.5 \cdot 10^{-4}$ & 59 \\
    $1.25$    &   $-0.5$       &   $1.9 \cdot 10^{4}$    & $5.5 \cdot 10^{8}$   &  $2.6 \cdot 10^{-2}$  &  $7.0 \cdot 10^{-3}$ & $9.6 \cdot 10^{-5}$ & $1.5 \cdot 10^2$\\
    $2.5$     &   $-1$    &  $3.8 \cdot 10^{4}$  &  $2.2 \cdot 10^{9}$  & $1.7 \cdot 10^{-1}$  &   $2.7 \cdot 10^{-3}$     & $6.0 \cdot 10^{-5}$ & $3.9 \cdot 10^2$\\
    $5$     &   $-2$    &   $7.6 \cdot 10^{4}$    &  $8.8 \cdot 10^{9}$  &  $1.1$  &  $1.0 \cdot 10^{-3}$      & $3.7 \cdot 10^{-5}$ & $1.0 \cdot 10^3$ \\
    $10$     &   $-4$    &  $1.5 \cdot 10^{5}$     &  $3.5 \cdot 10^{10}$  & $7.6$  &  $4.1 \cdot 10^{-4}$ & $2.3 \cdot 10^{-5}$ & $2.5 \cdot 10^3$ \\ \hline
\end{tabular}
\caption{Flow parameters for the used experiments: values for $\epsilon$ are obtained from simulations of the experimental setup \cite{Ostilla2014,Ezeta2018}.}
\label{tab:flow parameters}
\end{table*}
\\
The flow contains turbulent Taylor vortices, as reported by van der Veen et al. \cite{vdVeen2016}. The turbulent Taylor vortices are present throughout the measurement volume, where the flow is axisymmetric around the vertical axis.\\
To properly image the white chiral particles inside the setup, the inner cylinder was spray-painted black, and LED-strips were added around the top and bottom of the outer cylinder to provide sufficient light and contrast. The chiral particles were tracked during the experiment using 4 high-speed cameras: 4 Photron Mini AX-200 cameras equipped with a \unit{50}{mm} lens, imaging \unit{1024}{px} $\times$ \unit{1024}{px}, at a resolution of approximately \unit{135}{\micro \meter \per px}. The cameras image a small volume (\unit{8}{cm} $\times$ \unit{3}{cm} $\times$ \unit{15}{cm}) centred approximately at mid-height of the cylinder, which was visible on all 4 cameras simultaneously. Figures \ref{fig:Setup_schematic}(a) and \ref{fig:Setup_schematic}(b) show the positioning of the cameras and the measurement volume, respectively.
\begin{figure*}
    \centering
    \includegraphics[width=0.8\linewidth]{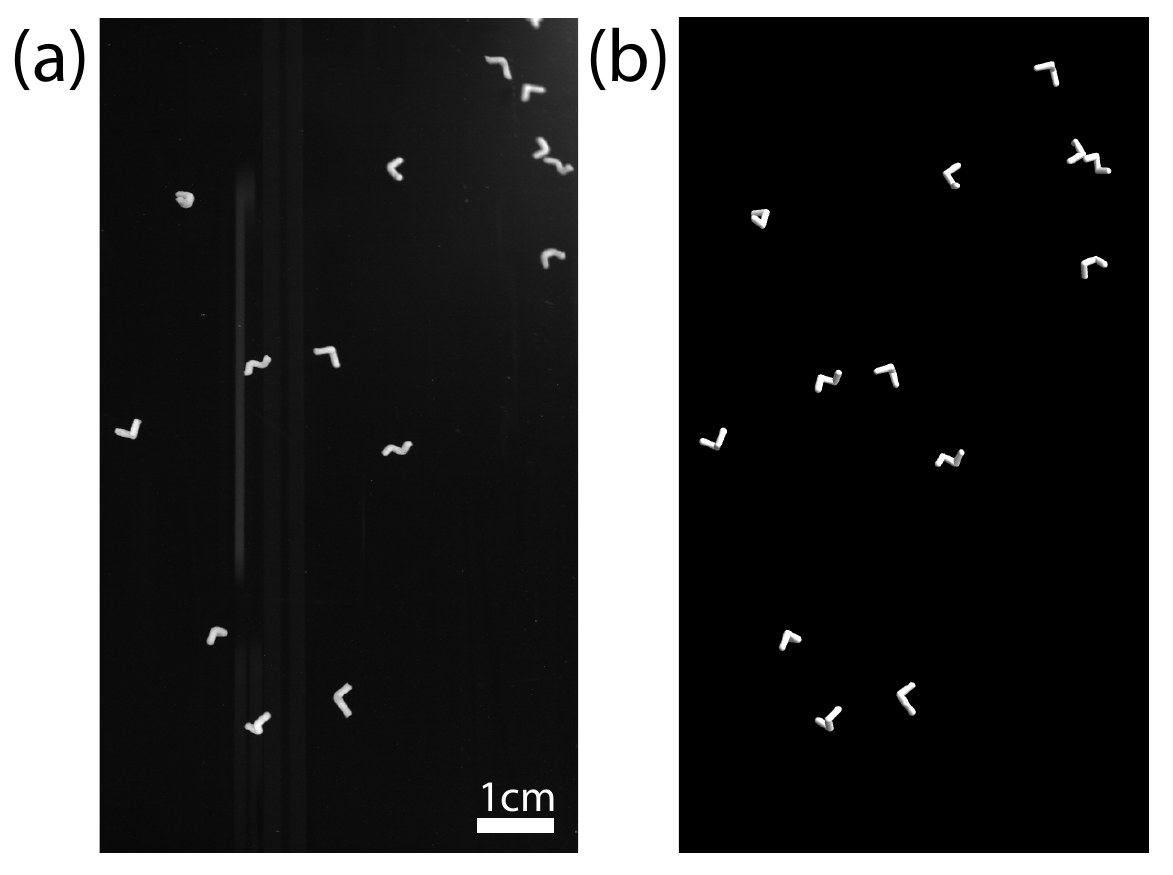}
    \caption{(a) A typical image captured by a high-speed camera during a measurement. Chiral particles are visible as grey shapes. Image was taken for a measurement with both particle chiralities at $\text{Re} = 3.8 \cdot 10^4$. (b) Reconstructed 3D chiral particles, constructed using tracking methods described in previous work \cite{Flapper2024}. An animation of a raw recording alongside the particle reconstruction is included in the supplementary material.}
    \label{fig:Particle_reconstruction}
\end{figure*}
\\ Figure \ref{fig:Particle_reconstruction}(a) shows a raw image of the chiral particles recorded by one of the cameras during a measurement at $\text{Re} = 3.8 \cdot 10^4$. These imaged chiral particles were reconstructed, giving the 3D reconstruction shown in Figure \ref{fig:Particle_reconstruction}(b). These reconstructed particles were found by tracking the location and orientation of the imaged particles, as described in previous work \cite{Flapper2024}. The main principles of the tracking method are described below.

\subsection{Particle tracking}\label{subsec: part track}
To prepare the 3D particle tracking, we calibrated the high-speed cameras using a triangular target which spans the gap between the cylinders, and we rotated the target in the azimuthal direction. This calibration provided the camera parameters for all used cameras. During the experiments, the chiral particles were recorded by all 4 cameras, enabling the location and orientation tracking. Tracking the location of the particles was done by finding the centroid of all detected particles, and centroid-matching the particles using a method of choice (in this case we used a ray-traversal method by Bourgoin and Huisman \cite{Bourgoin2020}). An important caveat for these particles is that the centroid of the imaged particle does not correspond to the centre of mass of the chiral particles, in fact, the centre of mass is outside the particle itself. To find the centre of mass of the particle, we recalculated the position of the centre of mass after finding the orientation, as described in previous work \cite{Flapper2024}. Finding the orientation is done by finding the images of a single particle on all four cameras. We create a synthetic particle with a known orientation, and compute the projections of the synthetic particle as seen by the cameras. These synthetic projections are compared to the experimentally found images, and we compute the error between these experimental and synthetic images. The orientation of the synthetic particle is optimised using a Nelder--Mead algorithm to minimise the error between the synthetic and experimental images, thereby finding the correct orientation of the chiral particle. Figures \ref{fig:Particle_reconstruction}(a) and \ref{fig:Particle_reconstruction}(b) show a raw image of the chiral particles in the described Taylor--Couette setup and the 3D reconstructed chiral particles, respectively. One of the particles in the raw image is missing in the reconstruction, which is due to the particle being outside the measurement volume. This method of finding the orientation is performed for all recorded frames, yielding the orientation of many chiral particles over time. We note that the orientation tracking method does not work for overlapping particles, resulting in approximately $1\%$ of the imaged particles being discarded. As soon as these particles no longer overlap, the particles can be tracked again. An animation of a raw recording alongside the reconstructed particles can be found in the supplementary material.

\begin{figure*}
    \centering
    \includegraphics[width = 0.75 \linewidth]{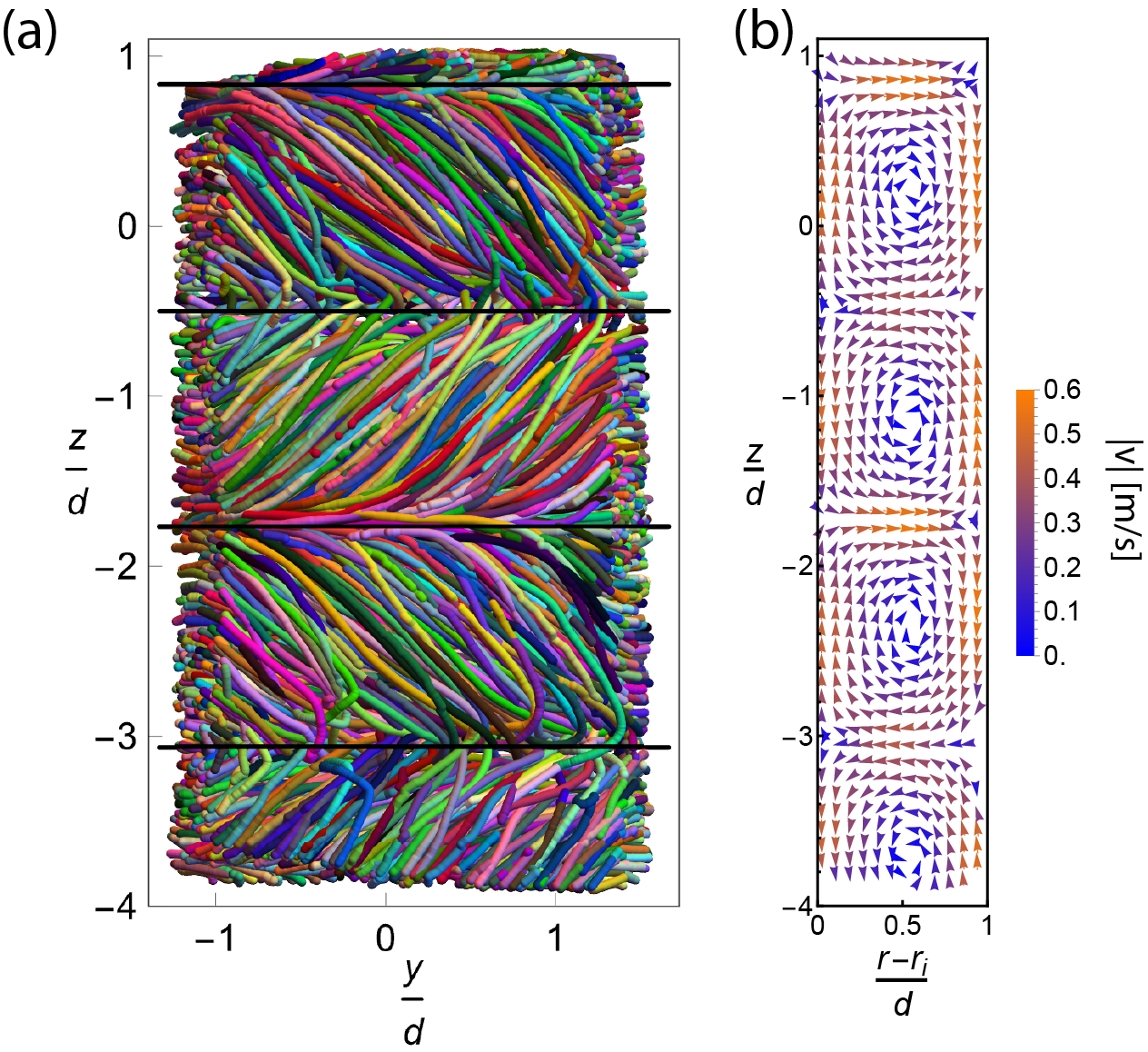}
    \caption{a) All particle location tracks for $\text{Re} = 1.5 \cdot 10^5$. The dominant azimuthal velocity is from left to right in this visualisation. Black lines are added for illustrating the boundaries of the vortices. The vertical dimension has its origin at the mid-height of the cylinder. The horizontal dimension is centred around the middle of the inner cylinder. b) A vector plot of the radial and axial velocity components of the particles, taken in a radial slice of the measurement volume, also for $\text{Re} = 1.5 \cdot 10^5$.}
    \label{fig:Tracks}
\end{figure*}
\section{Results} \label{sec:Results}
\subsection{Location data}
The tracking algorithm described above is used to track the location and orientation of the chiral particles over time. The particle location tracks are shown in Figure \ref{fig:Tracks}(a) for all tracked particles during a single measurement at $\text{Re} = 1.5 \cdot 10^5$.
Solid black lines are added to this figure to illustrate the boundaries of the turbulent Taylor vortices. These tracks clearly show the presence of the Taylor vortices and illustrate the alternating rotation directions of the vortices. From this data the vortex size is determined to be approximately \unit{4}{cm} in height: much larger than the typical particle size, which is \unit{5}{mm}. An animation of the tracks forming Taylor vortices can be found in the supplementary material.
The Taylor vortices are also visibly present in Figure \ref{fig:Tracks}(b), which shows the azimuthally averaged particle velocities for a measurement at $\text{Re} = 1.5 \cdot 10^5$. Here the particle locations over time are used to find the velocities of the particles by finite-differencing. This figure shows that the particle velocity in the radial-axial plane is lower in the vortex core, and higher at the edges of the vortices.
Given the clear Taylor vortices seen in Figure \ref{fig:Tracks}, the location data of a particle indicate whether a particle is in a clockwise or anticlockwise vortex. This can be used to find whether the particle handedness is under- or overrepresented in a clockwise or anticlockwise Taylor vortex, thereby indicating a separation by handedness. Counting the chiral particles in each vortex by chirality gives the results in Table \ref{tab: chirality tallies}, and shows no clear division between the two chiralities over the complete range of Reynolds numbers.
\begin{figure*}
    \centering
    \includegraphics[width = 0.7\linewidth]{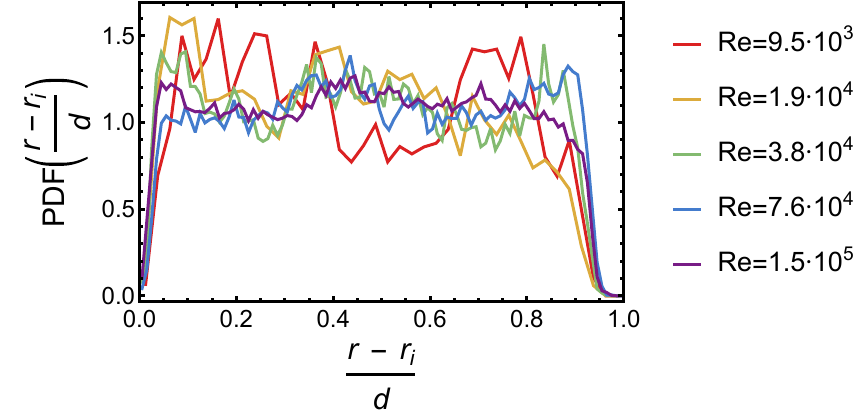}
    \caption{PDF of the radial distribution of particles for all measured Reynolds numbers. The radial distance is nondimensionalised, and the shape of the measurement volume is taken into account.}
    \label{fig:Radial_dist}
\end{figure*}
\begin{table*}[]
\begin{tabular}{|C{1.5cm}|C{1.3cm}|C{1.5cm} C{1.5cm}|C{1.6cm} C{1.6cm}|C{2cm} C{2cm}|}
\hline
Re & Tracks & \multicolumn{2}{c|}{Total tally} & \multicolumn{2}{c|}{Clockwise vortex tally} & \multicolumn{2}{c|}{Counterclockwise vortex tally} \\
   &        & Left                    & Right  & Left                          & Right       & Left                             & Right           \\ \hline
$9.5 \cdot 10^3$  & 442      & \multicolumn{1}{c|}{59\%}  & 41\%      & \multicolumn{1}{c|}{63\%}        & 37\%           & \multicolumn{1}{c|}{55\%}           & 45\%               \\ \hline
$1.9 \cdot 10^4$  & 962      & \multicolumn{1}{c|}{59\%}  & 41\%      & \multicolumn{1}{c|}{59\%}        & 41\%           & \multicolumn{1}{c|}{59\%}           & 41\%               \\ \hline
$3.8 \cdot 10^4$  & 1459      & \multicolumn{1}{c|}{58\%}  & 42\%      & \multicolumn{1}{c|}{56\%}        & 44\%           & \multicolumn{1}{c|}{60\%}           & 40\%               \\ \hline
$7.6 \cdot 10^4$  & 2525      & \multicolumn{1}{c|}{56\%}  & 44\%      & \multicolumn{1}{c|}{56\%}        & 44\%           & \multicolumn{1}{c|}{56\%}           & 44\%               \\ \hline
$1.5 \cdot 10^5$  & 1470      & \multicolumn{1}{c|}{62\%}  & 38\%      & \multicolumn{1}{c|}{62\%}        & 38\%           & \multicolumn{1}{c|}{62\%}           & 38\%               \\ \hline
\end{tabular}
\caption{This table shows the tallies of the left-handed and right-handed particles in the clockwise and anticlockwise vortices, and in total. The total number of tracks are also displayed to illustrate the size of the dataset.}
\label{tab: chirality tallies}
\end{table*}
Comparing the handedness of the particles within the vortices, we notice that this is not different from the overall ratio between the handedness of the particles. Therefore, the Taylor vortices are not observed to separate the chiral particles by handedness. The separation by chirality using a vortical flow as found at smaller scales \cite{Aristov2013} is therefore not observed at these measured Reynolds numbers. \\
The particles' location data is also used to investigate the radial distribution of the particles, which is shown in Figure \ref{fig:Radial_dist}. This figure shows that the particles are approximately homogeneously distributed over the radial distance between the inner and outer cylinder, where the shape of the measurement volume is taken into account. 
The particles do not show a clear preference for a certain radial distance, as is expected for density-matched particles. The lower number of particles close to the inner and outer cylinder are explained by the dimensions of the particle: the particle's size, shape and orientation prohibit the particle centre from getting very close to the inner and outer wall.
\begin{figure*}
    \centering
    \includegraphics[width=\linewidth]{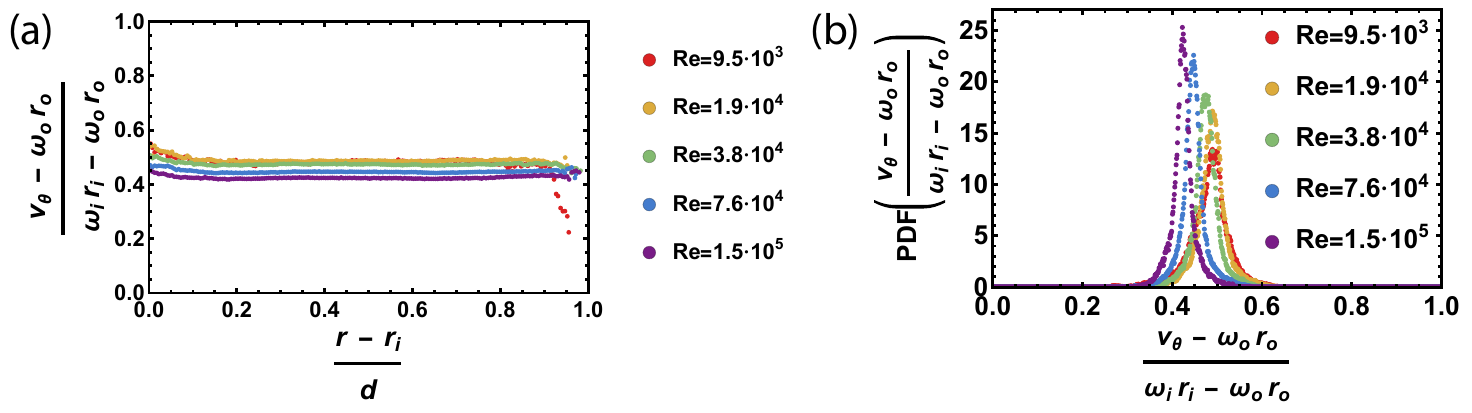}
    \caption{(a) Particle azimuthal velocity normalised by inner cylinder velocity as a function of dimensionless radial position. (b) PDFs of the dimensionless azimuthal velocities for all investigated Reynolds numbers.}
    \label{fig:Velocity_plots}
\end{figure*}
\\
Finally, Figure \ref{fig:Velocity_plots}(a) shows the azimuthal particle velocity normalised by the inner cylinder velocity as a function of the dimensionless radial position. Here we binned the dimensionless radial position in bins sized $0.005$, and took the average velocity for each bin. Figure \ref{fig:Velocity_plots}(b) shows the PDFs of the particle azimuthal velocities for the measured Reynolds numbers over all radii. The normalised particle azimuthal velocities decrease as the Reynolds number increases, which we attribute to the counter-rotation of the concentric Taylor--Couette cylinders. The shapes of the curves in Figure \ref{fig:Velocity_plots} seem similar across the Reynolds numbers, but no collapse of the data is found.\\ 
The highest investigated Reynolds numbers go into the ultimate turbulence regime \cite{Huisman2012}, for which the angular velocity of the flow normalised by inner cylinder angular velocity is approximately $\frac{\omega}{\omega_i} \approx 0.10$ for $a = 0.4$ \cite{Gils2012}. This corresponds to a value of approximately $\frac{v_\theta - \omega_o r_o}{\omega_i r_i - \omega_o r_o} \approx 0.43$, which we found at the highest investigated Reynolds numbers.

These results clearly show the chiral particles following the Taylor--Couette flow. Despite the particles' location closely adhering to the turbulent Taylor vortices, this provides no evidence of the orientation statistics of the chiral particles. Since finite-size fibers have a preferential orientation in similar turbulent Taylor--Couette flow \cite{Bakhuis2019}, we carefully dissect the orientation statistics to verify whether a preferential orientation also exists for chiral particles.

\subsection{Orientation data}
\begin{figure*}
    \centering
    \includegraphics[width = 0.8\textwidth]{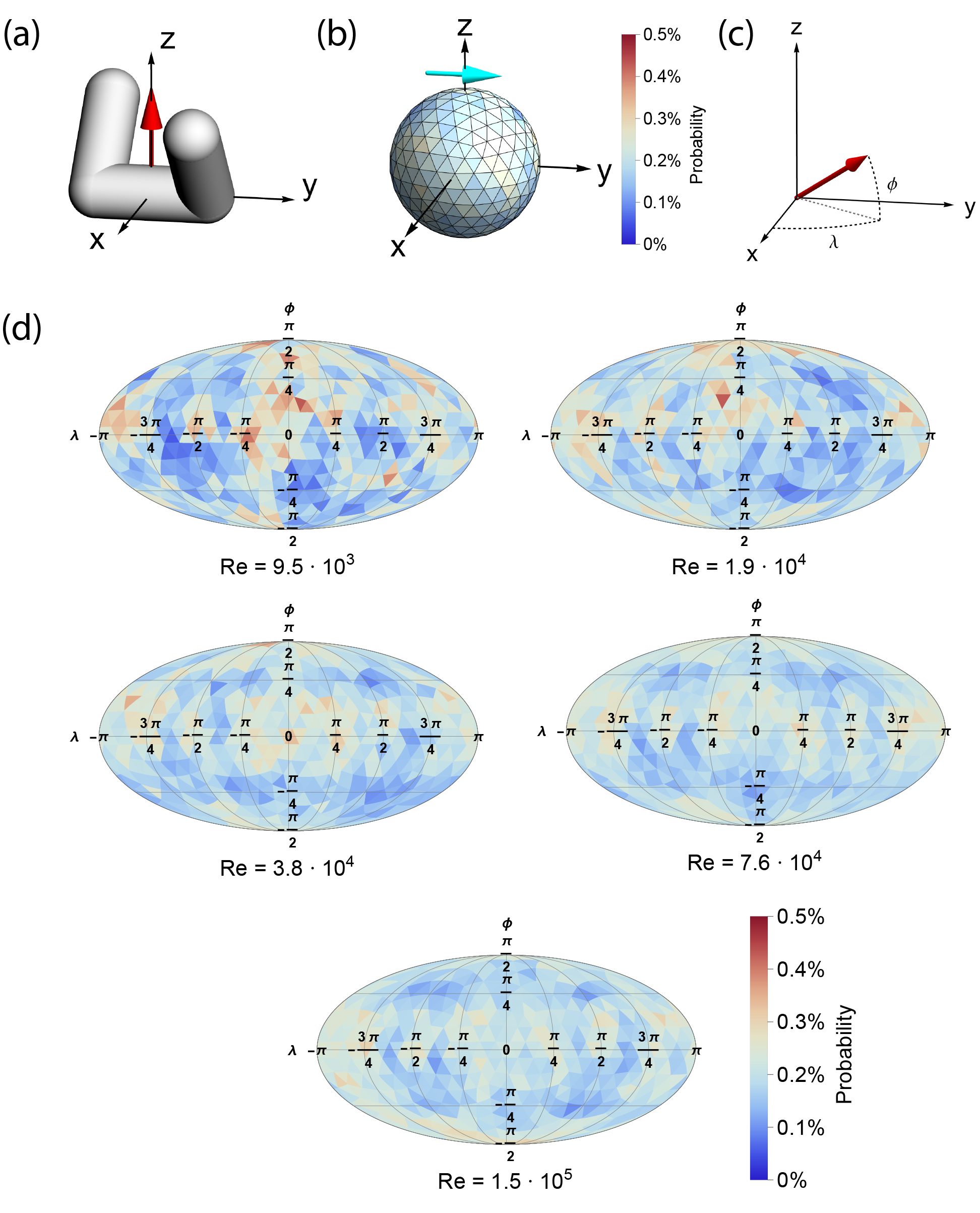}
    \caption{(a) Particle and the corresponding pointing vector (shown by the red vector), which is used to find the particle alignment. (b) Pointing vector density for $\text{Re} = 1.5 \cdot 10^5$ calculated as a percentage of the total number of orientation vectors found, where 1470 particles were tracked, and $1.5 \cdot 10^5$ vectors were computed in total. The cyan arrow indicates the flow direction. (c) The definition of the latitude and longitude coordinates. (d) Projection of the pointing vector density shown in (b), using the Mollweide projection.}
    \label{fig:Orientation_sphere}
\end{figure*}

Using the orientation reconstructions of the chiral particles over time, we can determine whether the particles have a preferential orientation, and whether the chiral particles align with the flow. We define a pointing vector for the chiral particle as shown by the red arrow in Figure \ref{fig:Orientation_sphere}(a), and trace its direction over time, where we compensate for the curvature of the experimental setup. The pointing vector density mapped onto a sphere for $\text{Re} = 1.5 \cdot 10^5$ is shown in Figure \ref{fig:Orientation_sphere}(b), where the colours of the triangles indicate the fraction of data where the pointing vector of the chiral particle points in the direction of the triangle. This projection shows the data for all detected particles over a complete measurement (for a single Reynolds number).
The light blue arrow indicates the direction of the dominant flow velocity (azimuthal). The sphere is divided into 500 triangles, where the orientation density is compensated by the solid angle for each triangle to ensure a fair comparison. Only particles away from the wall with radial position $0.25<\frac{r-r_i}{r_o - r_i}<0.75$ are included to prevent the wall from directly affecting the orientation. As previously noted, the sphere is divided into 500 triangles, meaning that a homogeneous distribution of the pointing vector would give a density of $0.2 \%$ for each triangle. Note that, for the following analysis, the rotation around the red pointing vector in Figure \ref{fig:Orientation_sphere}(a) is immaterial.\\
To show the pointing vector density in all directions, the sphere shown in Figure \ref{fig:Orientation_sphere}(b) is projected using the area-preserving Mollweide projection. Figure \ref{fig:Orientation_sphere}(c) shows the definition of the latitude $(\phi)$ and longitude $(\lambda)$ coordinates, used for the Mollweide projection. Figure \ref{fig:Orientation_sphere}(d-h) shows the projected pointing vector density for the measured Reynolds numbers, showing a uniform distribution (within the limited data we acquired) for each Reynolds number. 
Note that we have better convergence for the larger Reynolds numbers, as we have more data for those cases. There does not seem to be a general direction in which the pointing vector density is especially high or low. Again, the decreasing variation in pointing vector density for increasing Reynolds numbers can be attributed to the larger amount of data collected for higher Reynolds numbers. Therefore, the pointing vector densities show that there is no strong preferential orientation for chiral particles for these Reynolds numbers, unlike what Bakhuis et al. \cite{Bakhuis2019} found for finite-size fibers.
Similarly, we do not observe alignment of the pointing vector opposing the dominant particle velocity direction, which was observed when the particles settle in a quiescent fluid \cite{Piumini2024}.
\begin{figure*}
    \centering
    \includegraphics[width=0.75\linewidth]{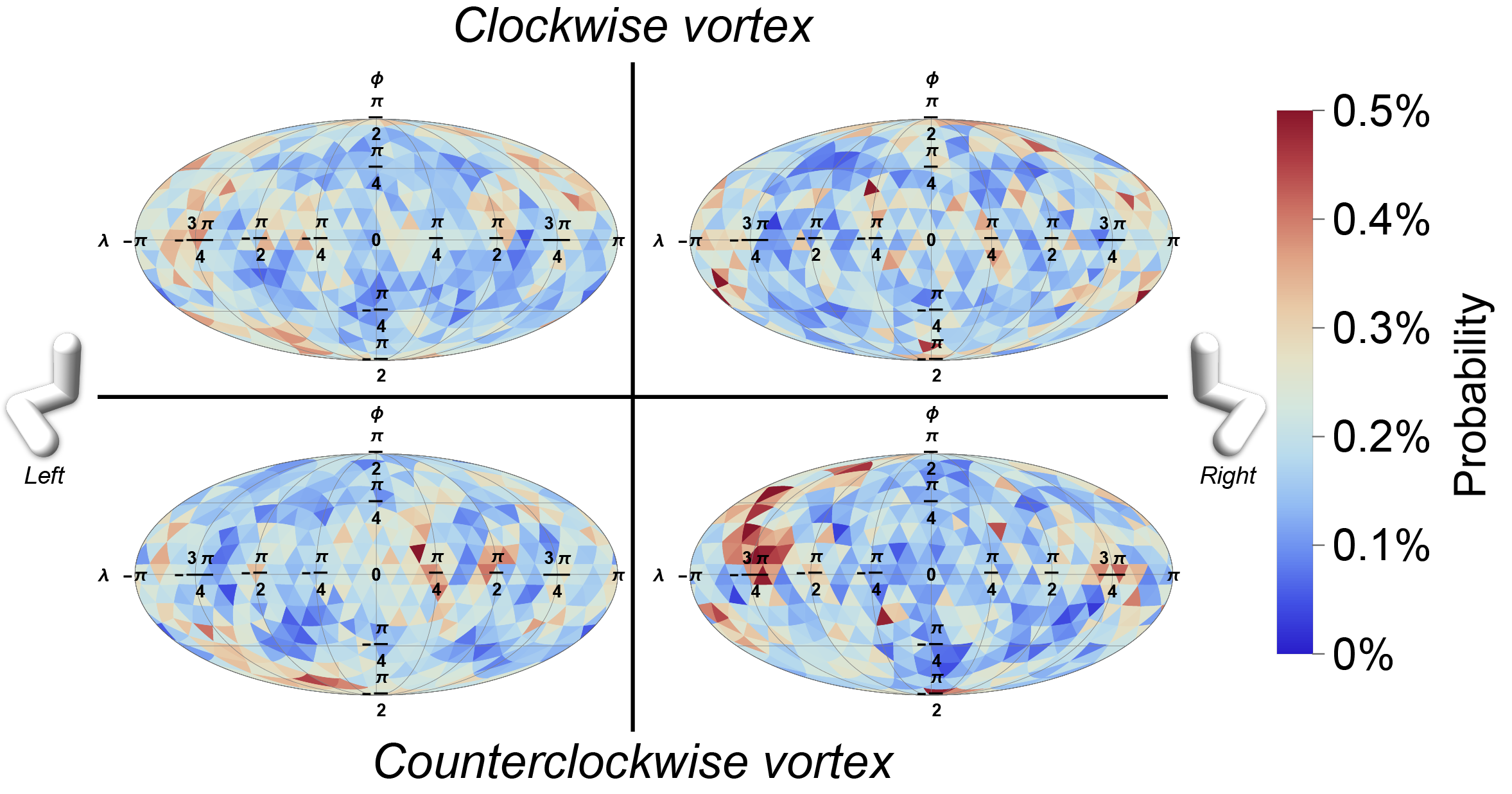}
    \caption{Mollweide projection of the pointing vector density for $\text{Re} = 1.5 \cdot 10^5$, split by particle chirality and vortex rotation direction.}
    \label{fig:Orientation_data_split}
\end{figure*}

\noindent
To check the uniformity of the data, they are split by particle chirality (left-handed and right-handed), and split between particles in the clockwise and counterclockwise vortices, giving a four-way split in the orientation data. The pointing vector density split by particle chirality and vortex rotation direction is shown in Figure \ref{fig:Orientation_data_split} for $\text{Re} = 1.5 \cdot 10^5$.
Again, we see no very high or very low density area in the plots, indicating no preferential orientation of the chiral particles. Comparing the different panels in the figure, no noteworthy differences are visible. Therefore, none of the observed cases show evidence of the particle orientation aligning in any direction. For all studied Reynolds numbers, this finding is the same, as demonstrated in Appendix \ref{app:orientation_rotation}, showing no preferential orientation when splitting the data as in Figure \ref{fig:Orientation_data_split}. This indicates no alignment of the chiral particles with the flow for the range of high Reynolds numbers investigated in this study. This is unlike fibers, which have a preferred alignment angle with respect to the Taylor--Couette inner cylinder, even at high Reynolds numbers \cite{Bakhuis2019}. Our findings for chiral particles are in agreement with numerical simulations, which showed that the orientation statistics are not affected by particle chirality at high Reynolds numbers \cite{Piumini2024}. Since no preferential orientation or alignment is found, we find no translation-rotation coupling, and no driving mechanism for the chiral particles to separate. Similarly, no distinct differences between the particle dynamics are expected given the random particle orientations.

\subsection{Rotation data}
\begin{figure*}[h]
    \centering
    \includegraphics[width = \textwidth]{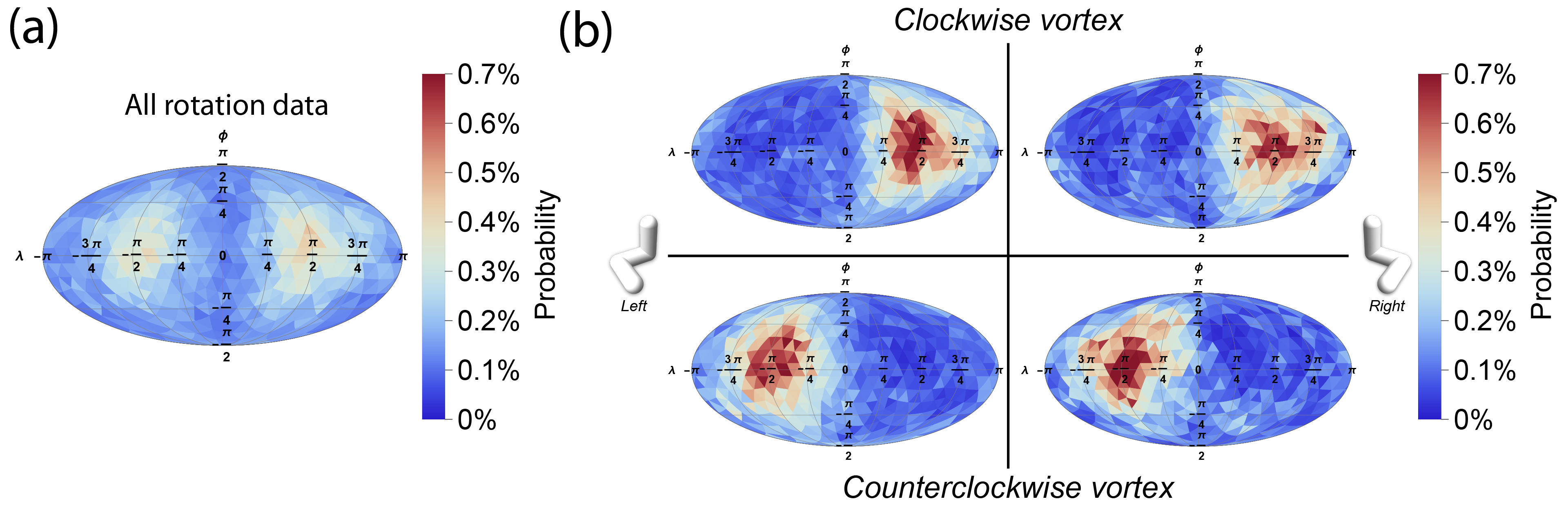}
    \caption{(a) The projected rotation vector density shown for $\text{Re}_i = 1.5 \cdot 10^5$. (b) The projected rotation vector densities split for the particle handedness, and whether this rotation occurs in the clockwise or counter clockwise vortex.}
    \label{fig:Rotation data split}
\end{figure*}

\begin{figure}
    \centering
    \includegraphics[width=\linewidth]{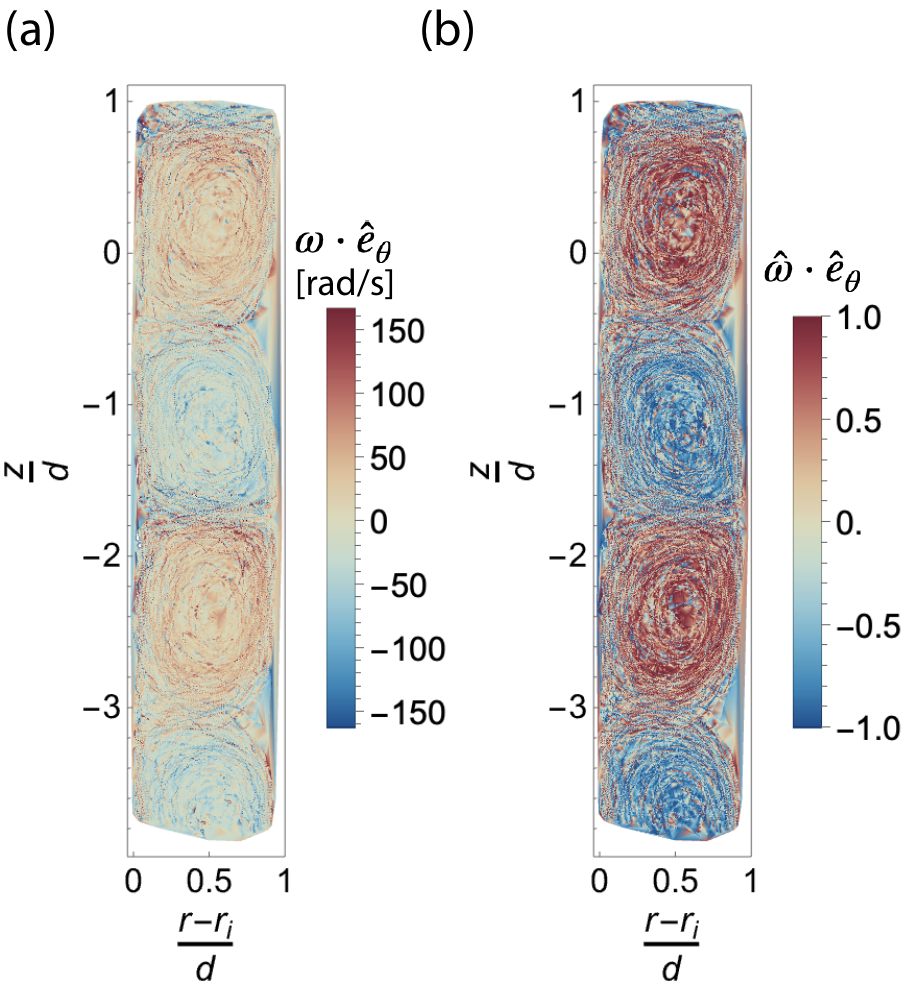}
    \caption{(a) Angular velocity of the tracked particles averaged over the azimuthal direction for $\text{Re} = 1.5 \cdot 10^5$. (b) Alignment of the angular velocity vector with the azimuthal direction for $\text{Re} = 1.5 \cdot 10^5$.}
    \label{fig:Angular_velocities}
\end{figure}

\begin{figure*}
    \centering
    \includegraphics[width=\linewidth]{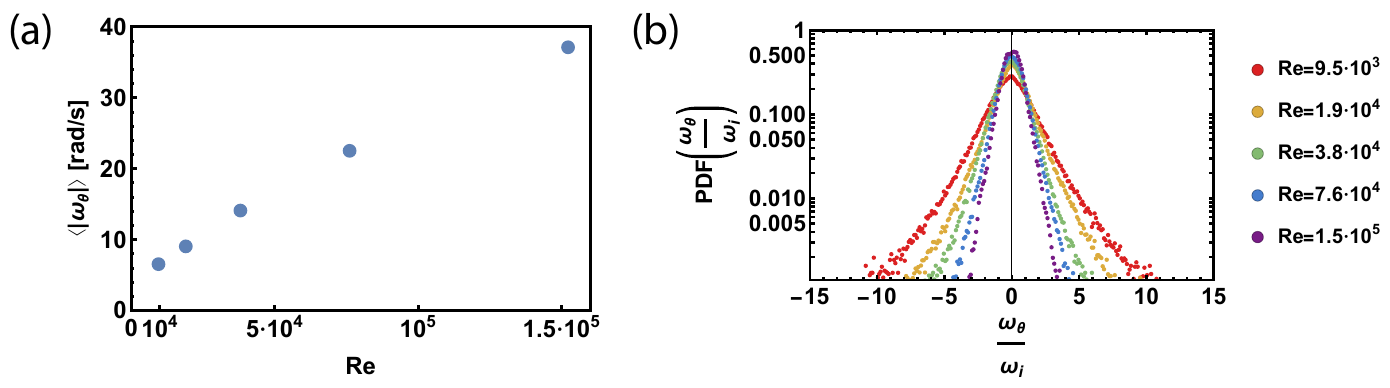}
    \caption{(a) Mean absolute angular velocity of the tracked particles for the investigated Reynolds numbers. (b) PDFs of the absolute angular velocities of the tracked particles.}
    \label{fig:Angular velocity plots}
\end{figure*}

From the orientation data over time, the particles' angular velocity $(\vec{\boldsymbol{{\omega}}})$ can be found\footnote{Note that the choice in the pointing vector (red arrow in Figure \ref{fig:Orientation_sphere}a) does not change $\vec{\omega}$.}. The rotation data is investigated in a similar way as the orientation data. The rotation vector is tracked over time, and a projection of the rotation vector density for all particles is shown in Figure \ref{fig:Rotation data split}(a). Note that this analysis investigates the direction of the rotation vector, and is unaffected by the magnitude of $\vec{\boldsymbol{\omega}}$.
The displayed rotation axis density shows two very clear high density areas, indicating a strong preferential rotation around two different axes. The direction of these axes correspond to the positive and negative azimuthal directions (the rotation directions of the different Taylor vortices). Once again, the rotation data can be split by particle chirality and vortex rotation direction, which is shown in Figure \ref{fig:Rotation data split}(b). Splitting this rotation data shows a clear difference between the vortices, but no notable difference between the particle chiralities. This trend holds over all investigated Reynolds numbers, which is explicitly shown in Appendix \ref{app:orientation_rotation}. We observe that the particles in the clockwise vortex rotate in a clockwise direction, whereas the particles in the counterclockwise vortex rotate in a counterclockwise direction. This demonstrates that the particle's rotation dynamics is dominated by the flow, and is not affected by the particle's chirality.\\
To further illustrate this point, the angular velocities are azimuthally averaged, which is shown in Figure \ref{fig:Angular_velocities}(a) for $\text{Re} = 1.5 \cdot 10^5$. 
This figure again shows the presence of strong Taylor vortices, and affirms that the particles follow the vortical flow. The different vortices show rotations in opposite directions, as is expected from Taylor vortices. Figure \ref{fig:Angular_velocities}(b) shows the dot product between the angular velocity vector of unit length and the azimuthal direction. This shows the strong alignment between the particle and vortex rotation direction, indicating that the particle's rotation dynamics are driven by the turbulent flow, and are unaffected by the particle chirality. Given that the particles closely follow the Taylor vortices, and have no preferential orientation (and therefore do not show any translation-rotation coupling), this result is expected, though nontrivial to predict a priori.\\
The magnitude of the particle angular velocity can be investigated further. Here one expects a higher angular velocity at higher Reynolds numbers, since the turbulent Taylor vortices are stronger, and the particle rotation has been shown to be dominated by the flow rotation. Figure \ref{fig:Angular velocity plots}(a) shows the average absolute angular velocities in the azimuthal direction for the measured Reynolds numbers. This figure shows that the average angular velocity indeed increases with the Reynolds number. The trend seems close to linear in the measured regime, though it should be noted that this seemingly linear relation between these quantities cannot exist over the entire range in the plot. Figure \ref{fig:Angular velocity plots}(b) shows the PDFs of the angular velocities in the azimuthal direction $\omega_\theta$ normalised by the angular velocity of the inner cylinder for all measured Reynolds numbers. We see that there is no general collapse, but the PDF becomes narrower as the Reynolds number increases, which we expect to be caused by the counter-rotation of the inner and outer cylinder.

Overall, the results clearly indicate that the strong turbulent flow overpowers any effects on the particle dynamics produced by particle chirality. These results conform to the findings by Piumini et al. \cite{Piumini2024}, which showed preferential particle dynamics at low Reynolds numbers, but isotropic particle dynamics as turbulence increased. While not entirely surprising, the multitude of relevant parameters makes it difficult to a priori predict this result for chiral particles in the Taylor--Couette flow.

One of the relevant parameters in this study is the particle size with respect to the Kolmogorov scale. For particles smaller than the Kolmogorov scale, velocity fluctuations only occur over scales larger than the particle-scale. As a result, small particles only experience a shear flow, and may therefore show periodic or preferential alignment, following Jeffery's equations \cite{Jeffery1922}. Such results have been found for fibers (even for those of finite size) in turbulent Taylor--Couette flow \cite{Bakhuis2019} for similar and higher Reynolds numbers ($8.3 \cdot 10^4 \leq \text{Re} \leq 2.5 \cdot 10^5$). We anticipate sub-Kolmogorov particles to follow the large scale structures of the flow, while being constantly reoriented due to the turbulence.

Larger particles (those larger than the Kolmogorov scale) have dynamics which are nontrivial to predict, and have been measured in this study. In this case, the particles are larger than the smallest turbulence features, meaning the fluid velocity is non-uniform over the particle-scale. Despite a particle therefore having a velocity difference with the surrounding fluid (on the small scales), this does not lead to a preferential orientation in this case. In the performed experiments, the particle Reynolds number $\text{Re}_\text{p}$ ranges over $220 \leq \text{Re}_\text{p} \leq 1900$, and is defined as:
\begin{equation}
	\text{Re}_\text{p} = \frac{\langle v_\theta \rangle \ d}{\nu},
\end{equation}
where we use the typical particle size $d = \unit{5}{mm}$. These high particle Reynolds numbers indicate the presence of a turbulent particle wake in the performed experiments. The presence of this turbulent wake is expected, and can also affect the particle dynamics: since the overall dynamics are consistent over all performed measurements, it is difficult to isolate the effect of the varying particle Reynolds number. Though in our case the particle-particle interactions through wake effects to be negligible due the low volume fraction of particles $(\phi = 1.7 \cdot 10^{-4})$.

Alongside the particle size and particle wake, there are many more parameters we expect to play a role in this study, such as the slip velocity or the particle-vortex size ratio. Therefore, estimating the dynamics of oddly-shaped particles is rather difficult. We expect the dynamics of left- and right-handed particles to differ at lower Reynolds numbers as found by Piumini et al. \cite{Piumini2024} and Aristov et al. \cite{Aristov2013}, though we cannot estimate the regime in which these differences occur.

\subsection{Temporal dynamics}
So far, the dynamics we have analysed can be obtained without regarding any temporal dynamics. To better understand the particle dynamics and what drives the observed dynamics, we find a governing timescale for the millimetric chiral particles in Taylor--Couette turbulence, and investigate how it scales with the Reynolds number in this regime of Taylor--Couette flow. To estimate the scaling of the timescale driving the particles, the particle velocity scaling is used, given by $v_d \propto (\epsilon d)^{1/3}$, where $d$ is the typical particle size. The typical particle timescale then scales as 
\begin{equation}
    \tau_d = \frac{d}{v_d} \propto \frac{d^{2/3}}{\epsilon^{1/3}}.
\end{equation} 
Previous work has found the scaling of the energy dissipation in Taylor--Couette as $\epsilon \propto \frac{\nu^3}{L^4} \text{Ta}^{1.4}$ \cite{Ezeta2018}. Here we note that $\text{Ta} \propto \text{Re}^2$, such that the expected particle timescale scales as
\begin{equation}
    \tau_d \propto \text{Re}^{-\frac{2.8}{3}}.
\end{equation}
Normalising this particle timescale by the Kolmogorov timescale (using the same energy dissipation scaling as before) then gives
\begin{equation}
    \frac{\tau_d}{\tau_\eta} \propto \frac{\text{Re}^{-0.93}}{\text{Re}^{-1.4}} \propto \text{Re}^{0.47},
    \label{eq:timescale_scaling}
\end{equation}
as the expected scaling for the normalised particle timescale with the Reynolds number. While not of fundamental importance, this scaling serves to verify whether the measured particle velocity is indeed caused by the turbulent velocity: when the measured velocities agree with the predicted scaling, this provides confirmation that the observed dynamics are driven by the turbulent flow.\\
Experimentally, the particle timescale is determined by computing the mean autocorrelation of the particles' azimuthal velocity (or angular velocity) for each Reynolds number. The first zero-crossing of the mean autocorrelation is taken as the particle timescale, which is then divided by the Kolmogorov timescale to create a dimensionless time.
\begin{figure}
    \centering
    \includegraphics[width=\linewidth]{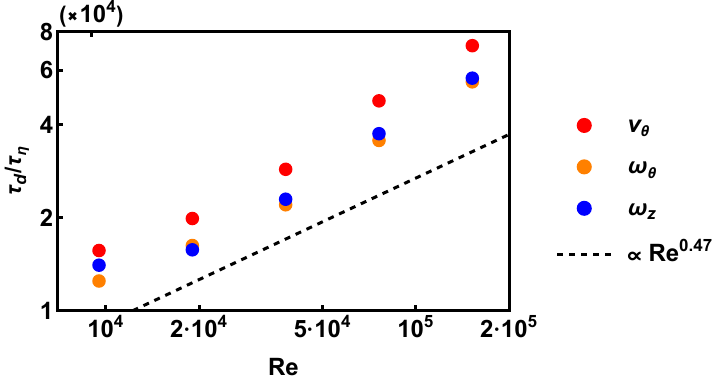}
    \caption{The particle timescale over the Kolmogorov timescale against the Reynolds number for multiple measured quantities. The dashed line shows the expected scaling.}
    \label{fig:particle_timescale}
\end{figure}
Figure \ref{fig:particle_timescale} shows how the normalised particle timescale scales with the Reynolds number. The dashed line shows the expected scaling for the normalised particle timescale, as found in equation \eqref{eq:timescale_scaling}. The predicted scaling gives a reasonable estimate for the scaling of the normalised particle timescale, governing the observed dynamics of the chiral particles in turbulent Taylor--Couette flow.

\section{Conclusions \& Outlook} \label{sec: Outlook & Conclusion}
In this work we reported on the dynamics of millimetric chiral particles in strongly sheared, turbulent Taylor--Couette flow containing Taylor vortices. The tracked particles closely follow the flow, where the particle locations over time trace out clear Taylor vortices. The particles are not separated by handedness between the alternating vortices: the particles are evenly distributed between the vortices, and are homogeneously distributed over the radial distance.\\
The chiral particles' orientation statistics show no preferential orientation or alignment to the flow, which we attribute to the strong turbulent flow. Any preferential alignment observed at lower Reynolds numbers is not observed at the measured Reynolds number range. Preferential orientation as found for fibers in turbulent Taylor--Couette flow is not found for chiral particles. Consequently, we also do not find any translation-rotation coupling since that would depend on the orientation, which seems to be uniformly distributed for our data. Splitting the data by particle chirality and flow vorticity shows that the handedness of the particle and flow do not affect the particle orientation. The rotation statistics are also flow-dominated, where the particle rotation direction coincides with the vortex rotation direction almost perfectly.\\
Any translation-rotation coupling shown at lower Reynolds number is therefore no longer discernible at these high Reynolds numbers. The effects of the particle chirality on its dynamics are completely overpowered by the turbulent flow, which dictates the particle dynamics in this regime ($9.5 \cdot 10^3 < \text{Re} < 1.5 \cdot 10^5 $). The observed particle dynamics are characterised by a timescale which is found to scale reasonably close to $\frac{\tau_d}{\tau_\eta} \propto \text{Re}^{0.47}$, which is consistent with scaling arguments.\\
Additional research is required to bridge the gap between the current research on high Reynolds numbers, and previous research on low Reynolds numbers, for example by Aristov et al. \cite{Aristov2013}. Investigating intermediate Reynolds numbers will reveal how the orientation shows up, and indicate the parameter range where particle chirality affects the particle dynamics.

\backmatter

\bmhead{Supplementary information}
Two supplementary videos are included. One video shows a recording of the chiral particles in Taylor--Couette turbulence, alongside a reconstruction of the tracked particles, both slowed down $33\times$. The second video shows the particle tracks over time (sped up $1.5\times$), visualising the Taylor vortices characteristic of Taylor--Couette turbulence. 

\bmhead{Data availability statement}
Data sets obtained and generated during the current study are available from the authors on reasonable request.

\bmhead{Acknowledgements}
The authors thank Federico Toschi for his comments and discussions. The authors thank Luuk Blaauw for performing preliminary experiments with chiral particles in the Taylor--Couette. We would also like to thank Gert-Wim Bruggert, Martin Bos, and Thomas Zijlstra for their technical support. This research has been funded by the Dutch Research Council (NWO) under grant OCENW.GROOT.2019.031. The authors report no conflict of interest.

\begin{appendices}

\section{Orientation and rotation data splits}\label{app:orientation_rotation}
\begin{figure*}
    \centering
    \includegraphics[width=\linewidth]{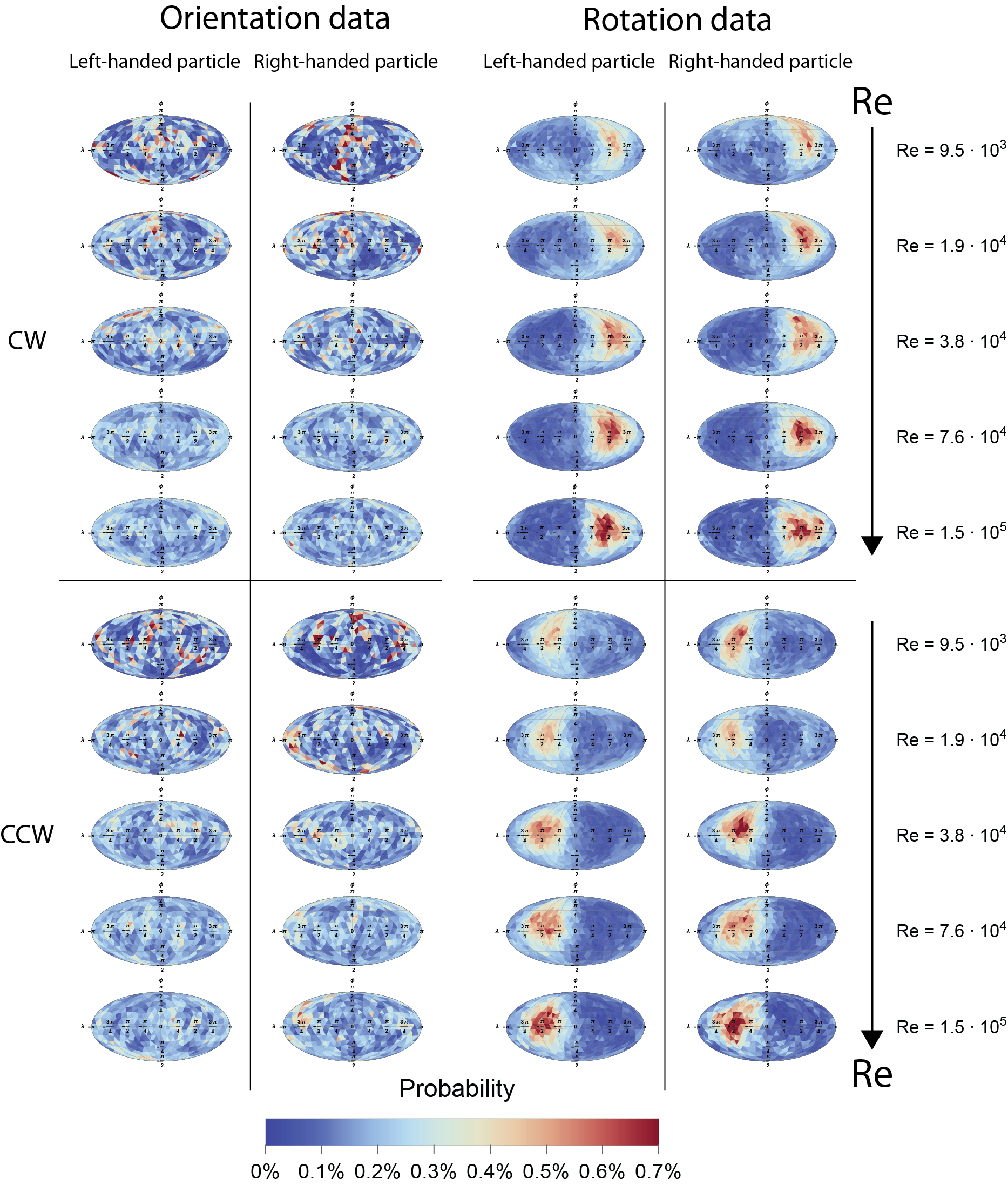}
    \caption{Orientation and rotation data split between vortex handedness and particle chirality, shown for all Reynolds numbers.}
    \label{fig:orientation_rotation_split}
\end{figure*}

The orientation and rotation data are split by particle chirality (left-handed and right-handed) and by vortex rotation direction (clockwise and counter-clockwise rotation) in Figure \ref{fig:orientation_rotation_split}, for all Reynolds numbers. Again, this affirms that the orientation data is randomly distributed, whereas the rotation direction is determined by the vortex.

\end{appendices}
\clearpage

\bibliography{sn-bibliography}

\end{document}